# Short-term cognitive fatigue of spatial selective attention after face-to-face conversations in virtual noisy environments

Ľuboš Hládek, Piotr Majdak, Robert Baumgartner

Acoustics Research Institute of the Austrian Academy of Sciences, Dominikanerbastei 16, 1010 Vienna, Austria

Corresponding author: Ľuboš Hládek – lubos.hladek@oeaw.ac.at

# Abstract

Spatial selective attention is an important asset for communication in cocktail party situations but may be compromised by short-term cognitive fatigue. Here we tested whether an effortful conversation in a highly ecological setting depletes task performance in an auditory spatial selective attention task. Young participants with normal hearing performed the task before and after (1) having a real dyadic face-to-face conversation on a free topic in a virtual reverberant room with simulated interfering conversations and background babble noise at 72 dB SPL for 30 minutes, (2) passively listening to the interfering conversations and babble noise, or (3) having the conversation in quiet. Self-reported perceived effort and fatigue increased after conversations in noise and passive listening relative to the reports after conversations in quiet. In contrast to our expectations, response times in the attention task decreased, rather than increased, after conversation in noise and accuracy did not change systematically in any of the conditions on the group level. Unexpectedly, we observed strong training effects between the individual sessions in our within-subject design even after one hour of training on a different day.

# Introduction

Communication in cocktail party situations, from the hearing-side perspective, requires both audibility and available cognitive resources (McGarrigle & Mattys, 2023; Ng et al., 2013; Pichora-Fuller et al., 2016; Sarampalis et al., 2009; Wang et al., 2025). While audibility aspects received substantial attention in research (Akeroyd & Whitmer, 2016; Best et al., 2017; Bischof et al., 2023; Bronkhorst, 2015), we have less understanding on how we use our cognitive resources in dealing with cocktail party listening, such as short-term memory, working memory or attention (Holt et al., 2022). When having a conversation, listeners use selective attention for keeping focus on an active speaker, divide their attention when listening to multiple sources, or switch the focus of attention between different sources (Choi et al., 2014; Shinn-Cunningham & Best, 2008). However, the ability to allocate attention may be affected by the inherent limits of cognitive resources leading to temporary cognitive fatigue, the feelings of tiredness or weariness after a cognitively demanding task. People with hearing loss experience such situations particularly often (Holman et al., 2019), which stresses the clinical importance for studying cognitive-related aspects of hearing.

Dual-task paradigms have been extensively used to assess the limits of cognitive processing during cocktail-party listening (Sarampalis et al., 2009). In these situations, the cognitive abilities may be taxed by performing a visual detection task (secondary task) during listening to speech sentences (primary task). In Sarampalis et al. (2009) people could keep speech intelligibility in the primary task due to additional noise, but the noise (in primary task) led to a decrease in performance in the secondary task. Using this technique is an effective way of assessing the effects of noise, or other types of sound, on concurrent cognitive tasks. However, this model would not work for testing cognitive effects in real conversations where people interact with each other (Beechey et al., 2019; Hadley et al., 2019; Hládek & Seeber, 2023). In context of testing assistive hearing devices, it is important to preserve natural motion behavior, as well as the context of the conversation, as much as possible without interference (Hendrikse et al., 2019; Hládek et al., 2019; Slomianka et al., 2024).

One possibility for testing cognitive processing in real face-to-face communication scenarios is to test cognitive after-effects such as ego depletion (Baumeister et al., 1998; Garrison et al., 2019). Similarly to studies on effortful listening, ego depletion studies assume that self-regulation is a cognitively demanding task due to limited cognitive resources, as proposed in the capacity model of effortful listening (Pichora-Fuller et al., 2016). The idea of ego depletion is generally in line with the studies on effortful listening (Alfandari et al., 2023; Zekveld et al., 2010) that showed that effortful listening leads to temporary depletion of cognitive resources, which was measured by decreases in arousal after effortful listening tasks, and by the effects of motivation on task performance and physiological measures (Richter, 2016).

Alfandari et al. (2023) hypothesized that the effect of effortful listening persists for a short time after an effortful task. They probed listening effort in terms of change of speech intelligibility and change of pupil diameter (PD) right before and after a 98-trial-long speech-in-noise 'load' task (40 min. long). They manipulated the effects of task difficulty by changing the signal-to-noise ratio (SNR) and task motivation by providing various monetary incentives. The results showed that the baseline PD decreased in conditions with increased task-induced listening effort, which they explained as diminished arousal related to mental fatigue. The higher monetary incentive led to an increased baseline PD in the high-SNR condition, whereas lower monetary incentives did not change the PD, indicating that motivation modulates the allocation of expected cognitive capacity. However, speech intelligibility measured at threshold in the probe task was unaffected, suggesting that 1) listening effort leads to short-term task-induced fatigue, which is persistent after effortful listening at the scale of minutes (the probe task took approximately 10 minutes), and 2) the fatigue might be assessed with standard methodologies like pupillometry. Unfortunately, that study did not assess the effect of fatigue on selective attention or other cognitive processing.

In two other studies (Gustafson et al., 2018; Key et al., 2017) researchers measured fatigue in children using a pre-/post-paradigm. They assessed arousal by the P300 magnitude in an oddball paradigm, by self-reports, and behaviourally by means of reaction times and lapses (trials in which the response time exceeded a threshold) in psychomotor vigilance task. They induced fatigue by a battery of speech processing tasks that were delivered over three hours and observed a decrease in arousal in the oddball paradigm, change in self-reports, and increase in reaction times and number of lapses indicated the effect of fatigue on attention.

Studies assessing the aftereffects of effortful listening (Alfandari et al., 2023; Gustafson et al., 2018; Key et al., 2017) focused on inducing fatigue with artificial laboratory tasks, but they did not probe the increase of fatigue in a realistic communication task. An important aspect of cocktail party communication is the use of spatial selective attention, and this aspect has not been considered yet when testing the sensitivity to short term-cognitive fatigue.

Our study addresses the hypothesis that short-term cognitive fatigue causes a temporary depletion of cognitive resources for spatial selective attention. In contrast to previous approaches, we measured the decrease in attention right after a realistic face-to-face conversation taking place in a virtual acoustic environment with reverberation, interfering conversations, and background noise. Spatial selective attention was measured before and after the conversation using an established syllable streaming paradigm (Deng et al., 2019).

# Methods

## Participants

Seventeen people were recruited from the "Vienna CogSciHub: Study Participant Platform" (Bock et al., 2014) and using personal connections at work. The targeted number of participants was obtained from the power analysis of based on within-subject variance from the previous experiment of the spatial selective attention task (Deng et al., 2019) using G*Power software (Faul et al., 2007).

Our participants (median age of 26 years, mean absolute deviation of 4.4 years, range from 22 to 35 years, five identified themselves as males and the rest identified themselves as females) were screened for hearing loss, defined as hearing thresholds of more than 20 dB HL on any frequency using pure-tone audiometry for audiological frequencies from 0.25 to 12 kHz. The participants were compensated by € 12.50 per session.

## Virtual acoustic environment (VAE)

The system consisted of open headphones, motion tracking, close directional microphone headsets, all of this controlled by rtSOFE (v 1.0) (Seeber et al., 2010; Seeber & Clapp, 2017; Seeber & Wang, 2021), a software for real-time acoustic simulations of the Munich underground station Theresienstrasse (Hládek et al., 2021; van de Par et al., 2022). This system created a virtual cocktail party situation for a real conversation between two participants. The personal microphone recorded the participants' voices, which were processed to add reverberation to the speech, therefore we created a complete acoustic simulation.

We used real-time simulations of the early parts of the binaural room impulse responses (BRIR), that were used to create the simulation. The early parts were obtained from the geometrical method implemented in rtSOFE, which implements the image source method and updates the room acoustics in real-time according to the motion of the participants in the laboratory. The motion data were obtained from the motion tracking (HTC Vive Pro 2.0) units installed on the wall of the lab and the units attached to the headphones of the participants. The headphones, microphones and the motion tracking units were connected by cables to wall outlets. Motivated by an earlier study (Hládek et al., 2021), the simulated BRIRs were 100 ms long. The rest of the impulse responses were obtained from publicly available binaural recordings of the Munich Underground (Hladek & Seeber, 2022). Details are summarized in Table 1. The simulated BRIRs (early parts) were convolved using direct convolution with the input signals to reduce latency. The late part was partitioned into multiple blocks and convolved using frequency domain convolution. The convolution engine provided the output in real-time without clicks or artifacts, which was fed to the headphones.

During the experiment, we used open headphones (AKG K1000) which did not block the direct sound from the other conversational partner. The headphones played simulation of the reverberation of one own voice, reverberation of the partner's voice, interfering talkers, and background noise. They were connected to an AD/DA converter (Hammerfall DSP, RME, Germany) via one of the two headphone amplifiers. The gain of the amplifiers was set prior to the experiment during the headphone calibration procedure, and the pre-amps were secured for accidental change of gain. Additionally, personal headset microphones (AKG MicroMic C 520 Vocal) were installed on each participant. Together with a reference microphone, that was used for calibration, they were routed to the same AD/DA converter via microphone pre-amps. The AD/DA converter collected all inputs. Two almost identical setups were used to provide real-time simulations for two participants. The computers were connected by a SPDIF cable to exchange audio channels with a minimal delay. Since the room acoustic simulation was computationally expensive, rtSOFE ran on multiple computers connected by local network. We created a set of custom scripts and commands that started and stopped all programs on multiple computers and whole procedure using one command.

The headphones had motion tracking units (HTC, Vive Pro 2.0) attached to them. The units were connected to a separate PC with a USB cable. The units provided positional and rotational information and sent them to the room acoustic simulation via custom Python scripts. The motion tracking system was calibrated using room setup procedure of the Vive system. The room calibration was performed routinely before experiments. The update rate of the motion tracking was 45 Hz which was limited by the available computational power for rtSOFE simulations (the motion tracker could provide up to 90 Hz). Each update from the motion tracking unit triggered an update of room acoustics. Thus, the room acoustic simulation had the same update rate. The positions of the participants were related to the midpoint of the dedicated experimental space. The virtual space was referenced to the Position 1 of the Underground scene (Hladek & Seeber, 2022; van de Par et al., 2022) marked on the floor of the laboratory and with additional marks on the floor of the experiment room we defined orientation of the acoustic model in the virtual space.

We used a set of equalization and calibration filters to calibrate levels and equalize the output across frequencies. After convolution of the input signals with BRIRs, the input signals were convolved again with the headphone equalization filter. This filter accounted for the coloration of the headphones and was computed from headphone transfer function. This function was measured for each headphone using measurement steps described in Majdak et al. (2007) and (2010). The resulting filter was obtained by averaging five measurements obtained from repeated measurements with the headphones repositioned. The averaging was done in the frequency domain on a logarithmic scale to account for the broadening of auditory filters in higher frequencies. Level calibration of the filter was performed with the calibrated sound level meter (B&K 2260) and artificial ear (B&K 4154) using a 1 kHz test signal (Brüel&Kjær Type 4231). The full-scale signal was calibrated to 94 dB SPL.

Input signals from the personal microphones were calibrated before the experiment. The participant was asked to speak for 20 s into the fitted microphone and a calibrated omnidirectional measurement microphone (PreSonus PRM1) placed at a 1 m distance and pointing to the participant. We created a Matlab tool that created equalization FIR filters by comparing spectra and levels of the outputs of the two microphones. The resulting FIR filter has the length of the simulation impulse response (256 taps). Due to specifics of rtSOFE implementation, the microphone equalization filter had to be applied into the source directivity simulation in the rtSOFE software. The calibration procedures of headphones, microphones and the motion tracking units that were created for this project were validated in a test that involved the whole processing chain (Götzke et al., 2025)

The room simulation software rtSOFE allowed: defining source directivity patterns, environmental settings (humidity, speed of sound), simulation settings (order of simulation), and mapping of the omnidirectional room impulse response to the desired reproduction method (binaural or loudspeaker reproduction). In the present study, the sound sources were simulated with the directivity of human voice (Flanagan, 1960), early reflections were simulated up to the third order, and the simulated room impulse responses were mapped to a non-individualized BRIR stored in SOFA format (Majdak et al., 2007, 2010). We use the option for real-time BRIR interpolation in rtSOFE to achieve smooth transitions between positional updates. The reverberation of conversing partners' voices was simulated without direct sound since people heard themselves directly during conversations. Other virtual sources were simulated with direct sound and reverberation.

## The acoustic scene

The acoustic scene for the conversations included two participants of the experiment and additional speech distractors with intelligible speech and background babble noise. The speech distractors were close-mic recordings of two people having a conversation in controlled acoustic environment (i.e., recordings of conversational speech between two people without reverberation and interference). The recordings were obtained from the database KEC (Arnold & Tomaschek, 2016) and two people spatialized to Positions (add) and additional two people were spatialized to Positions (add). Hence, the participants of the present experiment stood surrounded by 4 virtual people who were chatting on ad hoc topics, whose acoustic position was constat since the movement of the participants was considered in rtSOFE simulation. Additionally, the scene included simulation of background babble noise that was mapped to the fixed positions (without position update) which was created by convolution of many speech signals (from the same database) with positional impulse responses distributed around the center of the acoustic scene (note that the recording positions in the underground scene were in a circle around Position 1). In result participants an acoustically realistic simulation of environment where they could hear out individual voices that update position with movement of participants, hear reverberation of own and co-locutor's voice and hear noisy background babble. The sound level was 72 dBA, which was obtained as an average sound level across about 2 minutes of scene recording from the headphones (B&K 2260). However, substantial moment-to-moment fluctuations were present in the output signal.

During the recording of KEC database, two people were standing in two different sound booths and communicated on a freely chosen topic in German language for 30 minutes. The recordings were of high quality and did not include cross talk. Original recordings included parts with artificially inserted broadband noise to mask potentially sensitive information. These parts were cut-out from the recordings, and the missing parts were crossfaded with 200 ms linear ramps. Although, this made artificial gaps in sentences, the noise bursts (in the KEC database) would have been too disturbing.

## The conversation

During the conversation, the task of the participants was to talk to each other. Participants received the following instructions from the experimenter: "Imagine that the two of you were standing next to each other at the train station full of people. Now, it is a great opportunity to get to know each other just by having a conversation for about 30 minutes. During this time, you will hear conversations of other people, but you will not see them. You can safely ignore what they say by focusing on your partner. The task for the current experiment is just to keep talking and keep the conversation running even if it is difficult due to the background noise. You can speak freely to each other. However, keep in mind that some topics might be uncomfortable for the other person. For instance, topics like religion, politics, other people's appearance, or future, may not be appropriate for these 30 minutes. Choose a small-talk topic of your choice such as books you read, movies you watched, or places you visited. Or talk about your hobbies and activities you like. An example of a good topic is hearing loss as a global and personal problem. Also remember that if for whatever reason you do not feel comfortable, you can leave the experiment at any time without giving a reason. There is no penalty for doing that."

The conversation block started with fitting both participants with headphones and microphones. The next step was to calibrate personal microphones by speaking to the personal microphone (fitted to the person) and reference microphone placed at predefined distance at the same time. The conversation started and ended with the playback of the background sounds and added reverberation to own voices. The experimenter helped with the procedure and then left the testing room (IAC Acoustics, 120-A-Series) and started the experiment. After the conversation block, the experimenter helped participants to switch to the next part of the experiment.

## Spatial Selective Attention Task (SSAT)

The experiment used SSAT used previously (Deng et al., 2019) to assess attentional fatigue before and after conversations. The paradigm was almost identical to the HRTF condition (non-individualized) of the previous study expect we used 160 trials instead of 150. In comparison with the previous study, we used another set of HRTFs but identical syllable recordings. The stimuli of the SSAT were binaural mixtures of two syllable streams. Each of the two streams was made of three syllables from a set of /ba/, /da/, and /ga/ with replacement. Each syllable was of 0.4 s duration, and it was recorded by a high-pitched male (F0 = 125 Hz), a low-pitched male voice (F0=91 Hz), or a female voice (F0 = 189 Hz). The streams were presented over headphones (type) routed through a headphone amplifier (type) connected to a soundcard (type) near another experimental PC. The stimuli were spatialized using non-individualized HRFTs at +-30 degrees eccentricity, such that one stream was perceived to the left and one to the right. Two streams were formed in a way that one stream had male and the other female voice. The same syllable was never played simultaneously from the left and right, but the same syllable could have been played one or more times within one stream. The onsets of the first syllables from the two streams were time aligned, whereas the onsets of the second and the third syllables were offset. To minimize temporal cues and enhance spatial cues, one stream was called a leading stream, the other lagging. In the leading stream, the onsets of the fist and the second and the second and the third syllable were separated by 400 ms. In the lagging stream, the onset of the fist and the second syllable was offset by 600 ms and the second and the third was separated by 400 ms.

The task was to identify three target stream syllables of a binaural mixture. People always head a series of two syllable streams over headphones in a sound attenuated booth and provided answers using a keyboard. One syllable stream was a target stream, and the other one a distractor stream. The target stream was spoken by a female or a high-pitched male voice, the distractor was spoken by a talker with opposite sex. The sex of the target talker was determined pseudo randomly with equal probability on every trial.

The SSAT was split into trials. While participants were sitting in front of a computer screen in sound attenuated booth, they wore headphones and initiated the experiment. Trials started with a fixation dot on the computer screen in front of the sitting participant. The participant was instructed to fixate the dot. The fixation dot was on for 1.2 s, after which an auditory cue for target direction was presented. The cue was syllable /ba/ spoken by a low-pitched male voice (F0 = 91 Hz) at target location. The target direction was determined pseudo randomly with equal probabilities for the left and right. The leading and lagging streams were also distributed equally for both sides. The distribution of the male and female talkers were distributed equally across sides. The low-pitched voice was not used for the target talker. The target and distractor streams were played simultaneously 0.8 s after the onset of the target cue. Text 'Response' appeared on the screen after the streams were over together with a green dot. Participants could type three numbers, which are labeled as syllables, using a keyboard while the dot was green. The color of the dot changed after 1.5 s to red, after which the keyboard inputs were not considered. Feedback was provided after each trial. The whole procedure consisted of 160 trials presented in four block and took approximately 17 minutes.

## Self-reported questionnaire

We used an adapted version of the NASA Task Load Index (TLX) (Hart & Staveland, 1988) to assess mental effort, fatigue, comfort and relationship with the co-locutor. Specifically, the participants answered these questions that were written on a sheet of paper by writing the responses: 1) How mentally demanding was to maintain the conversation with the other participant and specifically maintain attention at the other participant due to the presence of competing talkers and background noise? (100-point scale from very low effort to very high effort). 2) How much mental fatigue did you feel after the conversation/listening to noise? 1 (no fatigue at all) to 100 (extremely mentally tired). 3) How well have you known the person before this experimental session? (100-point scale from very little (1 - complete stranger) to very much (100 – the closest person in life)) 4) How comfortable was it to be in the noisy environment on scale from 1 (not comfortable) to 100 (very comfortable)?

## Procedures

The experiment had four sessions that were usually conducted on separate days. Maximum two sessions could be performed on one day but then the participant took at least 30-minute break between the sessions. On the first session, participants performed the audiogram, and training of the SSAT task. The training procedure was identical to the SSAT task in the main experiment, except that participant performed 12 runs of 40 trials instead of four in the main task. Before the procedure, the experimenter explained the procedure and stayed with the participant in the booth for a few trials to ensure that they understood the task. Afterwards, the participants continued independently. On average the training procedure took almost 1 hour but the participants were encouraged to take longer breaks at the end of the runs when they felt fatigued.

In the three following sessions participants performed the conversation, preceded and followed by the SSAT task, with another fellow participant in one of three conditions – always one condition per day. The conditions were: CONV_NOISE, CONV_QUIET, LISTEN_ONLY. In CONV_NOISE, participants had a conversation and were embedded in the virtual acoustic scene that was noisy and reverberant. In CONV_QUIET, participants had a conversation, but the virtual acoustic scene was switched off, thus, there was neither reverberation nor noise. In LISTEN_ONLY, participants stood quietly in the virtual scene for 30 minutes (alone or in pairs).

When repeating over the conversations, the co-locutor was always a different person. The order of conditions was pseudo randomized by the experimenter to accommodate for the availability of the participants and targeting for a different order of conditions for each person.

Each session started with the SSAT task in one of the laboratory booths.  Two participants performed the task in two booths simultaneously and then came to another large booth where they had the conversation. The conversation started with fitting and calibrating the microphones and fitting headphones with motion tracking units. Motion tracking units were calibrated occasionally but the data output was monitored instantly. After 30 minutes of the conversation, the experimenter entered the large booth, and helped the participants to move to the small listening booths for the post-testing of the SSAT task (same booths as for the pre-test). The change of the rooms and starting the procedure took about two minutes. After the SSAT post-test participants filled out the questionnaire. The whole procedure for each session took approximately 1.5 hours.

## Analysis

The data were collected using Matlab (Natick, MA) and Psychtoolbox (PTB3) (Kleiner et al., 2007), the outcomes of the SSAT tests and questionaries were analyzed in RStudio (Posit team, 2025), plotting was done using package ggplot2 (Wickham, 2016). We analyzed the accuracy of responding the in SSAT task and the time it took to complete all responses in each trial (the time when they inserted the 3rd syllable of the SSAT task), since our experimental script unintentionally recorded only this value.

Accuracies were averaged across all trials and three syllables. Response times were log-transformed and pooled across all trials and correspond to time of typing-in all three syllables. Statistical significance was assessed by paired Wilcoxon signed-rank test with Bonferroni-Holm correction

# Results

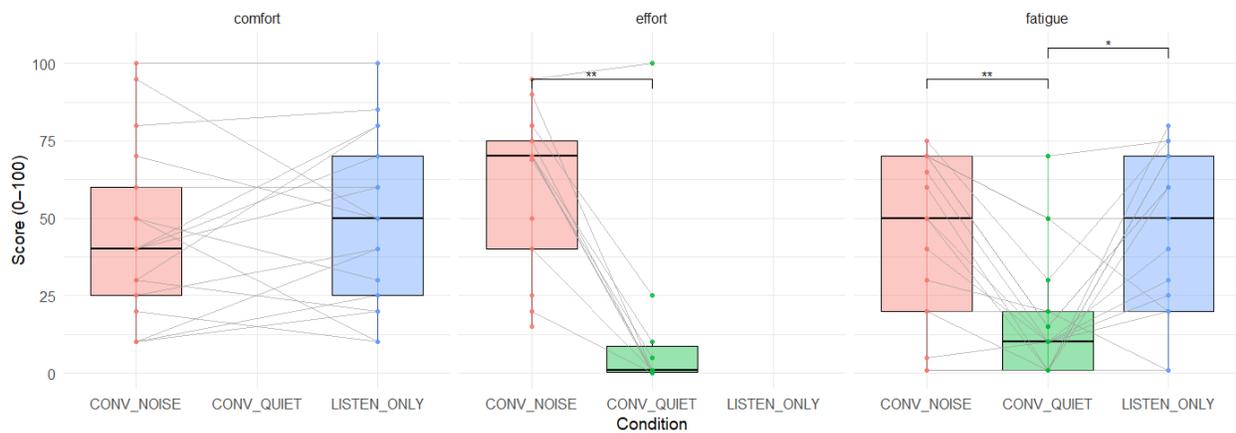

*Figure 1 Ratings of comfort, effort, and fatigue across intervention conditions. The ratings of comfort in CONV_QUIET were often not reported by the participants, the effort question asked about the conversation thus the answers in the LISTEN_ONLY were irrelevant because there was no conversation. Gray lines connect the ratings of individual participants. Significance levels: * p<0.05, ** p<0.01.*

Figure 1 shows self-reported ratings of perceived comfort, effort and fatigue of the experimental interventions. In some cases, participants did not provide answers. Statistical analysis revealed no change in comfort ratings. The effort ratings for the CONV_NOISE condition were significantly higher than for the CONV_QUIET condition (V(17,10) = 54, p = 0.004, r = 0.854). The fatigue ratings after the CONV_NOISE and LISTEN_ONLY conditions were significantly higher than after the CONV_QUIET condition (V(17,17) = 90, p = 0.006, r = 0.771; V(17,17) = 10.5, p = 0.014, r = 0.657).

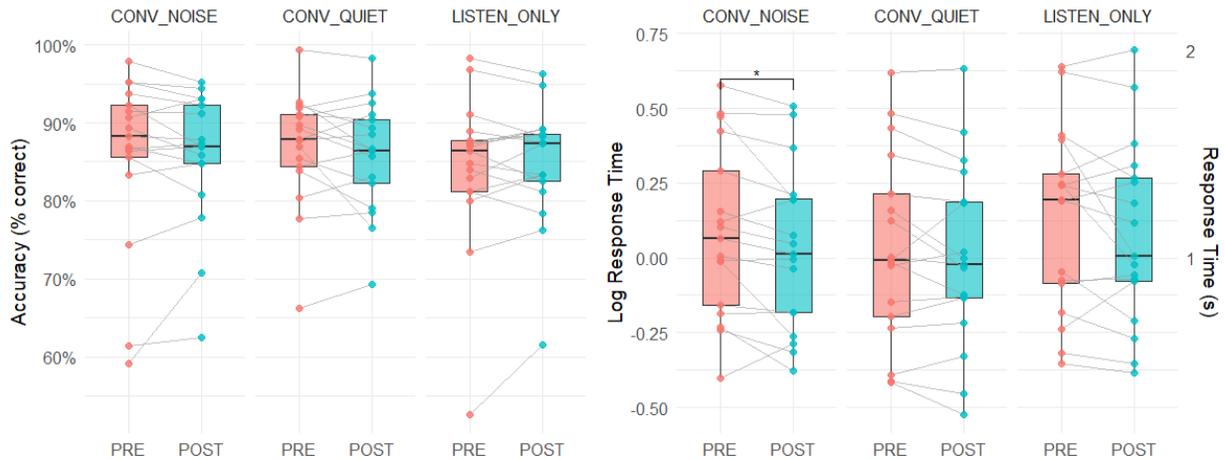

*Figure 2 Change of accuracy (left) and response times (right) between pre-test (PRE) and post-test (POST) for the three conditions. The accuracy was averaged across three syllables. The right panel shows logarithm of response time (left y-axis) and actual response times in seconds (right y-axis) that correspond to the time participants typed-in all answers. Individual points correspond to participants. Data of individual participants from pre-test to post-test are connected by lines. * denotes significance at p<0.05.*

Figure 2 shows accuracies and response times for the three conditions, before (PRE) and after (POST) the interventions with substantial variability between participants. Accuracies did not significantly change from pre- to post-testing in neither of the three conditions . Response times became significantly faster in the CONV_NOISE condition only (V(17,17) = 125, p = 0.02, r = 0.556).

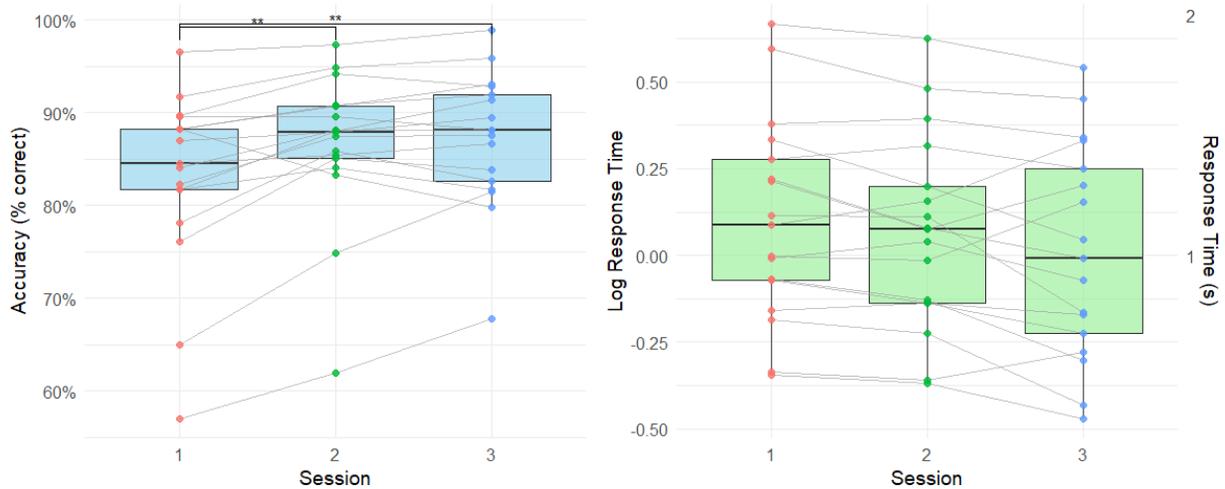

*Figure 3 Change in accuracy and response times across sessions, pooled across conditions and syllables. The ordinates of the right panel show the response time in a linear (right axis) and logarithmic (left axis) way. Data of individual participants are connected by lines. Significance levels: ** p<0.01.*

Figure 3 shows accuracies and response times as a function of session number with data pooled across conditions, trials and syllables. Response accuracies consistently improved between the first and final session (V(17,17)=11, p=0.003, r= 0.735) and second and third session (V(17,17)=14, p=0.006, r=0.689), while there was no significant change in log-response times between sessions.

# Discussion

The experiment tested whether spatial selective attention depleted after having an effortful communication task which was a free face-to-face conversation in noisy environment. Firstly, having a conversation was more subjectively demanding in noise than in quiet and the participants perceived more fatigue after a conversation in noise than in quiet. Contrary to our hypothesis, we observed faster responding in the post-test relative to pre-test and we did not observe a change in response accuracy after any of the intervention conditions (conversation in noise, conversation in quiet, passive listening). Thirdly, we observed a strong training effect from the first to the second session even though participants underwent a one-hour-long training session with 480 trials before.

## Effort and short-term cognitive fatigue after conversations

People perceived more subjective fatigue after the conversations in noise than after having conversations in quiet setting. Passive listening was more fatiguing than having a free conversation in quiet. Self-ratings of listening comfort did not systematically vary with the experimental conditions.

While we observed an increased effort and fatigue after a conversation in noise in comparison with a conversation in quiet, the objective behavioral measures of accuracy and response times in the SSAT task had different patterns with respect to our conditions. Similarly, the previous reports (Gustafson et al., 2018; Key et al., 2017) could not fully relate their objective measures to the subjective reports of short-term fatigue. The results confirm that the newly developed intervention – a free conversation in a noisy environment – leads to a substantially more subjective fatigue and perceived effort in comparison to the same conversation held in quiet. Interestingly, the passive listening task (LISTEN_ONLY) and active conversation (CONV_NOISE) obtained similar fatigue ratings, suggesting that the presence of noise was the determining factor of fatigue, not the conversation itself.

## Accuracy and response times of the SSAT task

The observed lack of accuracy differences is consistent with earlier observations for speech intelligibility ratings after an effortful task (Alfandari et al., 2023). We expected that the SSAT would be a more sensitive indicator of fatigue than the speech intelligibility task, but it well might be that similar processes were needed for solving the speech intelligibility task than the current

SSAT task. The cue that indicated target position minimized the spatial uncertainty which might have helped solving the task if fatigue decreased the attentional filter span (Hafter et al., 2007). Another aspect could be training effect from pre-test to post-test, which could have been more pronounced for participants with low initial scores. Three participants who started low in the CONV_NOISE systematically increased while this was not the case in other conditions.

We observed response times to decrease rather than increase, which is clearly at odds with the results on psychomotor vigilance tasks (Gustafson et al., 2018; Key et al., 2017). Yet, it is worth noting that those previous studies on listening fatigue used a behavioral task that tested sustained attention and used a three-hour-long session of effortful listening with multiple tasks, while the SSAT did not require sustained attention and our effortful conversation intervention lasted only 30 minutes.

## Training effect

The increase of accuracy without corresponding change in response times indicates a learning effect of the SSAT task. Previously, Laffere et al. (2020) observed increased detection sensitivity in a selective attention task after training on a previous day. They also observed better phase alignment after training on the selective attention task, which was measured by inter-trial phase coherence at 4 Hz of the EEG response, which was in coincidence with the stimuli presentation rate. Data of the previous experiment suggested an increase in attentional focus, however, it is unclear whether the current and the previous results related to procedural factors or whether some form of perceptual learning (De Larrea-Mancera et al., 2022) was taking place.

# Acknowledgements

Ľuboš Hládek is a recipient of the Seal of the Excellence Fellowship of the Austrian Academy of Sciences. Luca Götzke helped with the data collection and validation of the calibration procedures.